\documentclass[10pt,twocolumn]{article}

\usepackage[a4paper,top=1.8cm,bottom=1.9cm,left=1.7cm,right=1.7cm,columnsep=0.7cm]{geometry}
\usepackage[T1]{fontenc}
\usepackage[utf8]{inputenc}
\usepackage{lmodern}
\usepackage{microtype}
\usepackage{amsmath,amssymb,mathtools}
\usepackage{graphicx}
\usepackage{booktabs}
\usepackage{tabularx}
\newcolumntype{Y}{>{\raggedright\arraybackslash}X}
\usepackage{threeparttable}
\usepackage{array}
\usepackage{multirow}
\usepackage{siunitx}
\usepackage{xcolor}
\usepackage{enumitem}
\usepackage{caption}
\usepackage{subcaption}
\usepackage{float}
\usepackage{placeins}
\usepackage{flushend}
\usepackage{stfloats}
\usepackage{titlesec}
\usepackage{fancyhdr}
\usepackage{csquotes}
\usepackage[
  backend=biber,
  style=numeric-comp,
  sorting=none,
  sortcites=true,
  maxbibnames=99,
  doi=true,
  url=false,
  isbn=false,
  giveninits=true
]{biblatex}
\AtBeginBibliography{\raggedright}
\usepackage[colorlinks=true,allcolors=blue!55!black]{hyperref}
\usepackage[nameinlink,noabbrev]{cleveref}
\hypersetup{
  pdftitle={Has Scientific Talent Shifted from Depth to Breadth? Evidence across Papers, Knowledge Inputs, Careers, and Teams},
  pdfauthor={Xiaoshn Nee, Haobo Zhong, and Xiaomin Ni},
  pdfkeywords={science of science, research specialization, interdisciplinarity, scientific teams, artificial intelligence, OpenAlex, counterfactual forecasting}
}

\graphicspath{{figures/}}
\titleformat{\section}[hang]{\normalfont\Large\bfseries\filright}{\thesection}{0.8em}{}
\titleformat{\subsection}[hang]{\normalfont\large\bfseries\filright}{\thesubsection}{0.8em}{}
\titlespacing*{\section}{0pt}{1.8ex plus .4ex minus .2ex}{.9ex plus .2ex}
\titlespacing*{\subsection}{0pt}{1.4ex plus .3ex minus .2ex}{.7ex plus .2ex}
\setlist{nosep,leftmargin=*}

\newcommand{\runningtitle}{Scientific Talent from Depth to Breadth?}
\fancypagestyle{plain}{%
  \fancyhf{}
  \fancyfoot[C]{\thepage}
  }

\title{\textbf{Has Scientific Talent Shifted from Depth to Breadth? Evidence across Papers, Knowledge Inputs, Careers, and Teams}}
\author{%
Xiaoshn Nee$^{a}$\thanks{Corresponding author. Email \href{mailto:nixsh3@gmail.com}{nixsh3@gmail.com}}, Haobo Zhong$^{b}$, Xiaomin Ni$^{c}$\\[0.65em]
{\small $^{a}$Independent researcher}\\
{\small $^{b}$HSBC Business School, Peking University, Shenzhen City, Guangdong 518055, China}\\
{\small $^{c}$Artificial Intelligence Research Institute, Shenzhen University of Advanced Technology, Shenzhen, China}}
\date{13 September 2026}

\begin{document}
\raggedbottom
\maketitle

\begin{abstract}
Generative artificial intelligence raises a central question for scientific training and organization. Is research shifting from deep specialization toward broad individual knowledge? We examine this proposition across papers, cited knowledge, contributor histories, and teams using 47,959 articles from six fields over 2010--2025, 51,736 resolved cited works, and chronologically reconstructed prior publication histories for 1,754 randomly selected index contributors. From 2010 to 2022, team size increased by an estimated 37.3\% (95\% confidence interval [34.4\%, 40.3\%]), while paper topic breadth declined by 0.0144 on a 0--1 hierarchical distance scale. Cited knowledge was stable to modestly broader, revealing a divergence between focused outputs and the reach of knowledge inputs. Established contributors' prior breadth increased by 0.0190 [--0.0078, 0.0459] by 2019--2022, within a $\pm0.05$ equivalence bound assessed in sensitivity analysis. In mature citation windows, one standard deviation of focal depth was associated with 8.2\% higher $1+\mathrm{FWCI}$ [1.9\%, 14.9\%]; average breadth and interaction associations were smaller under the specified equivalence bounds. Post-2022 deviations from earlier trends were not systematic, and recent changes did not vary clearly with baseline AI intensity across 83 subfields. The findings support a differentiated structure of scientific expertise in which focused individual accumulation coexists with expanding collaboration and sustained access to diverse knowledge inputs.

\end{abstract}


\section{Introduction}

\begin{figure*}[!tbp]
\centering
\includegraphics[width=\textwidth]{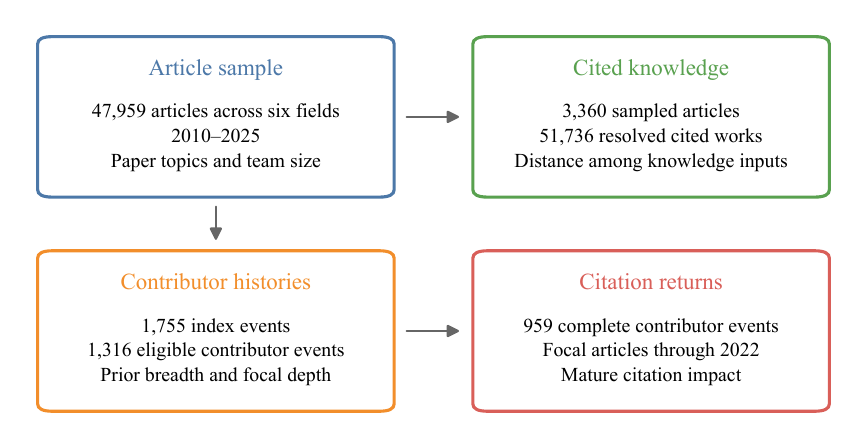}
\caption{Linked levels of scientific knowledge and capability. The article sample supplies two complementary branches, one following cited knowledge and the other reconstructing contributor histories. Citation returns link prior individual profiles to focal articles published through 2022, allowing the complete four year window used by OpenAlex's field weighted citation impact (FWCI). Arrows indicate data linkages. The reference branch includes 3,360 sampled articles, of which 3,215 meet the coverage criteria. The contributor branch contains 1,755 index events for 1,754 unique authors; 1,316 events meet the history and identity criteria, with 959 complete cases in the return model.}

\label{fig:design}
\end{figure*}

The growth of scientific knowledge makes it increasingly difficult for one person to master the components needed at the research frontier. The classic ``burden of knowledge'' account predicts longer training, narrower individual expertise, and greater reliance on collaboration as the stock of knowledge expands \parencite{jones2009}. Consistent with that account, teams have become dominant producers of scientific and technological knowledge \parencite{wuchty2007,fortunato2018}. Generative artificial intelligence (AI) changes how researchers navigate this constraint. A scientist can query literature, write and debug code, inspect data, and translate methods across fields through a single interface. This development motivates a consequential hypothesis that the scientific profile increasingly combines broad search, rapid learning, and coordination across tools and domains.

Understanding this transition requires connecting scientific talent to observable patterns of knowledge use. Publications reveal the expertise deployed in research, the collaboration through which it is mobilized, and the outputs that attract scientific recognition. Breadth can describe the topics represented in a paper, the knowledge sources it cites, the historical repertoire of an individual, or the expertise assembled through collaboration. Diversity itself contains variety, balance, and disparity \parencite{stirling2007}. Measures that combine these ingredients differently can yield contrasting conclusions about interdisciplinarity \parencite{zwanenburg2022,cantone2024}. A trend in scientific expertise is therefore best evaluated through several linked levels of observation.

The emerging literature on AI in science makes this distinction timely. Research using AI has been associated with higher individual productivity and impact alongside a contraction in the collective set of topics studied \parencite{hao2026}. Other large scale work finds heterogeneous associations with novelty and impact across knowledge spaces \parencite{bianchini2026}, while textual analyses document changes in scientific production associated with large language models \parencite{kusumegi2025}. AI also changes the relationship between access to information and scientific judgment \parencite{messeri2024}, motivating conceptual accounts of how expertise should be recalibrated \parencite{lin2026}. These findings invite a broader question about how the scientific system distributes knowledge across individual specialists, research outputs, and collaborative structures.

Prior research suggests that depth and breadth can contribute through different pathways. Unusual combinations can yield high impact science while increasing the risk of failure \parencite{uzzi2013,foster2015}. Interdisciplinarity has been associated with higher citation impact and lower publication productivity \parencite{leahey2017}, and the returns to distant interdisciplinarity vary across contexts \parencite{yegros2015,lariviere2015}. Specialization can also earn substantial and persistent rewards \parencite{rassenfosse2022}. The relative advantage of specialists and generalists depends on career stage and the pace of change at the knowledge frontier \parencite{teodoridis2019,ao2025}. These results motivate measuring depth and breadth separately and testing how their joint profile relates to scientific impact.

We develop a multilevel analysis using public bibliographic data. Figure~\ref{fig:design} connects articles to their cited knowledge and contributors' prior publication histories, then links individual capability profiles to mature citation outcomes. A general transition toward broader individuals would be supported by concordant changes in the knowledge combined in research, the repertoires of established contributors, and the returns to breadth or its complementarity with depth. We assess these implications over 2010--2025 and examine whether recent changes align with the diffusion of generative AI. Six primary tests were specified before estimating the capability and outcome models, with familywise error controlled using Holm's method. An independent reference sample, alternative measures, equivalence analyses, predictive counterfactuals, and continuous difference in differences provide complementary evidence.

The results identify a differentiated organization of knowledge. Team size expands while paper topics become more concentrated; cited inputs retain a wider reach, and accumulated focal depth is positively associated with mature citation impact. Established contributors' breadth changes more modestly. Our contribution is to quantify this separation across levels and show how it informs the debate over scientific talent. The emerging profile combines focused individual accumulation with collaborative capacity and access to diverse knowledge sources.

\section{Conceptual framework and tests}

\subsection{Observable structure of scientific expertise}

We use \emph{revealed capability structure} to describe the mix of depth, breadth, and coordination visible in scientific production. Three empirical patterns would support a broadening of this structure at the individual level. Papers would combine knowledge across greater intellectual distances, established contributors would approach focal problems with more diverse recent histories, and breadth would attract a positive return or strengthen the return to depth. Together, these patterns translate a claim about scientific talent into testable implications of deployed expertise.

Each layer captures a distinct part of the research process. Paper topic breadth describes the intellectual span of the output, while reference breadth records the distance among explicitly cited inputs. Individual breadth summarizes the distribution of a contributor's recent topics. Focal depth captures accumulated prior work in the subfield of the current article, and team size records the scale of collaboration. This separation allows locally specialized contributions to coexist with integration across distant sources, as suggested by earlier accounts of scientific interdisciplinarity \parencite{porter2009,evans2008}. It also distinguishes a change in individual repertoires from an organizational response through larger teams.

\subsection{Primary and triangulating tests}

The specified family contains six null hypotheses concerning the 2010--2022 trends in $\log(1+\text{team size})$ and paper topic breadth, the 2019--2022 changes in individual breadth and depth relative to 2010--2014, the interaction between breadth and depth in mature FWCI, and the mean post-2022 deviation from forecasts fitted to earlier years. The period 2019--2022 provides the final capability comparison before widespread generative AI access. Estimates for 2023--2025 characterize the more recent publication period, while mature citation models end in 2022.

Reference breadth provides an independent measurement of knowledge inputs. A pilot of 10 papers per field and year motivated a precision expansion to 35 papers per cell and at most 20 references per paper, selected using fixed random ranks based only on identifiers. The expansion was specified before collecting additional reference metadata. Pilot, expanded, clustered, and holdout estimates jointly assess sampling precision and stability. Table~\ref{tab:sample} summarizes how the linked samples correspond to the quantities of interest. The article sample estimates balanced trends across six fields, whereas the contributor sample represents a typical sampled article's randomly selected established contributor.

\begin{table*}[!t]
\centering
\caption{Study layers, analytic samples, and target quantities. Counts are after the exclusions shown unless otherwise stated.}
\label{tab:sample}
\begin{threeparttable}
\begin{tabularx}{\textwidth}{@{}p{2.8cm}rY Y@{}}
\toprule
Layer & $N$ & Inclusion and measurement & Target quantity \\
\midrule
Articles & 47,959 & Random sample of 500 OpenAlex articles in each of six fields and 16 publication years (2010--2025), excluding retracted and paratext records & Field adjusted trends in topic breadth and team size \\
Cited knowledge & 3,215 & Thirty-five papers per field and year with at least five references; up to 20 references sampled per paper; at least five and 70\% with a classified primary topic & Breadth of the knowledge inputs actually cited \\
Index contributors & 1,316 & One contributor selected uniformly from teams of 1--20 authors; 3--300 lifetime works; at least three prior works spanning two years; stable author identity & Revealed five year prior breadth and focal subfield depth \\
Citation returns & 959 & Model-complete index-contributor events published through 2022 with OpenAlex FWCI & Conditional citation associations with breadth, depth, and their interaction \\
AI intensity analysis & 83 subfields & At least 30 baseline papers (2010--2018) and 15 post-2022 papers; exposure fixed before the event window & Differential post-2022 change by prior AI intensity \\
\bottomrule
\end{tabularx}
\begin{tablenotes}[flushleft]\footnotesize
\item The article sample contains 48,000 records before exclusions. The cited knowledge expansion sampled 55,223 unique reference IDs and resolved 51,736 (93.7\%). The contributor models have different complete case counts; 1,316 is the eligible contributor pool.
\end{tablenotes}
\end{threeparttable}
\end{table*}

\section{Results}

\subsection{Expanding teams and diverging measures of breadth}

\begin{figure*}[!tbp]
\centering
\includegraphics[width=0.92\textwidth]{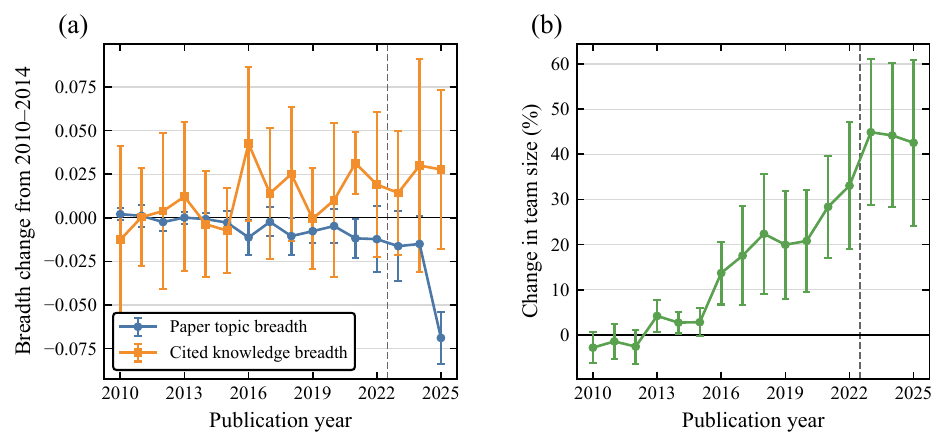}
\caption{Long term changes in breadth and team size. (a) Each breadth series is centered on its field specific 2010--2014 mean and then averaged across six fields; bars show 95\% intervals based on variation across fields. Topic breadth uses the primary exactly three topic metric. Cited knowledge breadth uses the expanded reference sample. (b) Team size is expressed as the percentage change from each field's 2010--2014 mean. Dashed lines mark the start of the post-2022 period.}
\label{fig:trends}
\end{figure*}

As shown in Figure~\ref{fig:trends}a, the distance among topics assigned to research outputs declined while the breadth of cited knowledge followed a flatter, mildly rising trajectory. Among papers with exactly three topics, the annual decline over 2010--2022 was 0.00120 per year (95\% CI [--0.00166, --0.00074], raw $p=3.17\times10^{-7}$, Holm $p=1.58\times10^{-6}$). The fitted cumulative change was --0.0144 [--0.0199, --0.0089] on the 0--1 hierarchical distance scale. This pattern persisted with uncertainty clustered within field and year cells (--0.00120 [--0.00190, --0.00050]), topic count controls, equal topic weights, maximum distance, and an indicator for crossing a broad field. Four fields had negative slopes, with computer science approximately flat and arts and humanities slightly positive. The pooled contraction therefore reflects a recurring pattern across distinct research settings. Restricting the article analysis to the contributor team size frame and giving each field and year cell equal weight also preserved a negative pooled slope, as detailed in Table~\ref{tab:s_selection}.

Figure~\ref{fig:trends}b places this increasing topical concentration alongside expanding collaboration. The specified annual coefficient for $\log(1+\text{team size})$ was 0.01738 [0.01602, 0.01875], with Holm $p<10^{-100}$. An auxiliary $\log(\text{team size})$ model among 38,285 pre-2023 articles with recorded authorships gives an annual coefficient of 0.02642 [0.02461, 0.02823], corresponding to a modelled increase of 37.3\% [34.4\%, 40.3\%] over 2010--2022. Its annual interval also remained positive with clustering within field and year cells [0.02172, 0.03112]. The plotted annual means summarize variation across fields, whereas these regression estimates quantify the common trend after field adjustment.

Standardizing recent papers to the earlier team size distribution clarifies the relationship between collaboration and topic concentration. Reweighting 2023--2025 papers to the 2010--2014 shares of single author teams and teams of 2--5, 6--20, and more than 20 authors retained a negative breadth change in all six fields. In engineering, the observed change of --0.0591 became --0.0367 after standardization; in medicine, --0.0420 became --0.0304. Team composition therefore accounts for part of the contraction in these fields, while substantial concentration also occurs within the standardized composition. The coexistence of larger teams and more focused output spans is consistent with finer coordination around specialized research problems.

The reference sample adds a complementary view of the knowledge used to produce those outputs. Among 2,619 eligible pre-2023 articles, cited knowledge breadth increased by 0.00228 per year, with an HC3 interval of [0.00002, 0.00454] ($p=0.048$). Clustering within field and year cells gave [--0.00053, 0.00509] ($p=0.112$), and the holdout excluding pilot articles yielded 0.00255 [0.00024, 0.00486] ($p=0.030$, HC3; $N=2,469$). Coarser measures based on the share of references outside the focal field and the effective number of cited fields were approximately stable. Taken together, the estimates support stable to modestly broader knowledge inputs alongside more concentrated output topics. This divergence is substantively useful because a focused contribution can continue to draw on intellectually distant sources. The measures share a topic hierarchy but aggregate different objects, so their trajectories describe complementary dimensions of research.

\begin{figure*}[!tbp]
\centering
\includegraphics[width=0.98\textwidth]{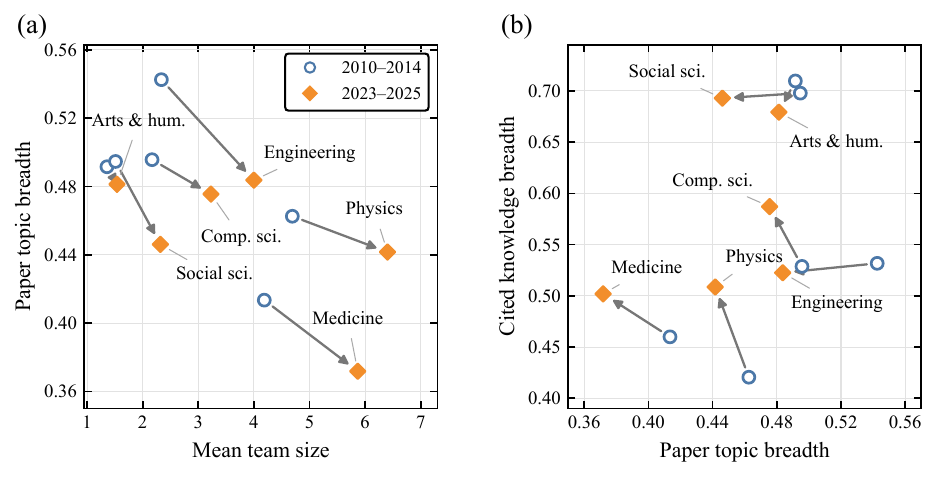}
\caption{Direct comparison of earlier and recent research profiles. Blue open circles show 2010--2014, orange diamonds show 2023--2025, and gray arrows connect the two means for each field. (a) Team size and paper topic breadth. (b) Paper topic breadth and cited knowledge breadth. Each coordinate averages the annual field means with equal year weights. Team size uses the full article sample, topic breadth uses articles with exactly three topics, and cited breadth uses the eligible reference sample. Field labels identify the recent endpoints.}
\label{fig:period_fields}
\end{figure*}

The paired positions in Figure~\ref{fig:period_fields} make this reorganization directly visible. All six fields move toward larger teams and lower paper topic breadth between 2010--2014 and 2023--2025. In engineering, mean team size rises from 2.34 to 4.00 as topic breadth falls from 0.543 to 0.484; medicine moves from 4.19 to 5.87 authors and from 0.413 to 0.372 in breadth. The knowledge input and output comparison in Figure~\ref{fig:period_fields}b adds a distinct pattern. Computer science, medicine, and physics move toward higher cited knowledge breadth while their output topics concentrate. Engineering and social sciences change little in cited breadth, whereas arts and humanities shows a more visible contraction in knowledge inputs. These paired positions connect a common organizational direction with variation in how fields combine knowledge inputs.

\subsection{Individual repertoires changed more modestly}

\begin{figure*}[!tbp]
\centering
\includegraphics[width=0.86\textwidth]{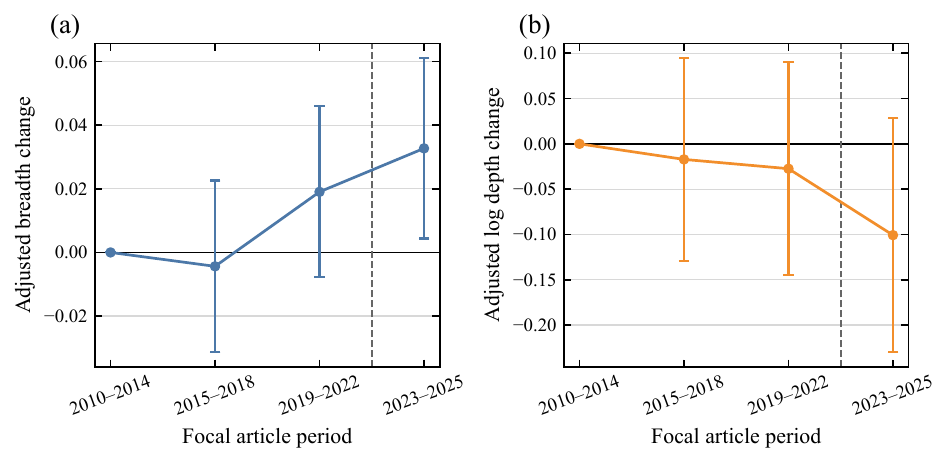}
\caption{Adjusted changes in individual capability. (a) Prior topic breadth. (b) Accumulated depth in the focal article's subfield, measured on a logarithmic scale. Points are weighted period contrasts relative to 2010--2014, with HC3 95\% confidence intervals. The baseline is fixed at zero. Both measures use publications strictly before the focal article, a five year lookback, and a five year exponential half-life. The dashed line separates the recent 2023--2025 period.}

\label{fig:individual}
\end{figure*}

Figure~\ref{fig:individual} traces adjusted changes in established contributors' prior breadth and focal depth. Controlling field, career age, and recent publication volume, five year breadth in 2019--2022 was 0.0190 higher than in 2010--2014 [--0.0078, 0.0459] ($p=0.165$, Holm $p=0.494$; model-complete $N=1,292$). Estimates from ten year histories and the last ten works were similar, while the ORCID subsample was centered near zero. The consistency across history definitions places the average change on a modest scale relative to the range of the breadth measure.

An equivalence analysis provides a direct assessment of magnitude. With a sensitivity bound of $\pm0.05$ on the 0--1 breadth scale, the two one sided test yielded $p_{\mathrm{TOST}}=0.012$ and a 90\% interval of [--0.0035, 0.0416]. This result supports a change within five hundredths of the scale under the specified model and bound. Focal depth changed by --0.0274 [--0.1451, 0.0903] in 2019--2022 ($p=0.648$, Holm $p=1.000$), leaving its direction less precisely resolved. Together, the panels place established individual profiles on a more gradual trajectory than the concurrent growth in team size.

The recent period adds a potentially informative shift in the point estimates. Five year breadth rose by 0.0327 [0.0044, 0.0610] in 2023--2025, compared with 0.0171 [--0.0114, 0.0456] using the last ten works and 0.0135 [--0.0205, 0.0474] among contributors linked to ORCID. Five year depth changed by --0.1008 [--0.2300, 0.0285], with more negative estimates under ten year and strict primary subfield definitions. The breadth estimates are positive across these recent specifications, while their attenuation under a fixed publication count highlights the role of how much recent output enters the repertoire. This pattern motivates following both the diversity and volume of individual research as the recent cohort matures.

\begin{figure*}[!tbp]
\centering
\includegraphics[width=0.894574\textwidth]{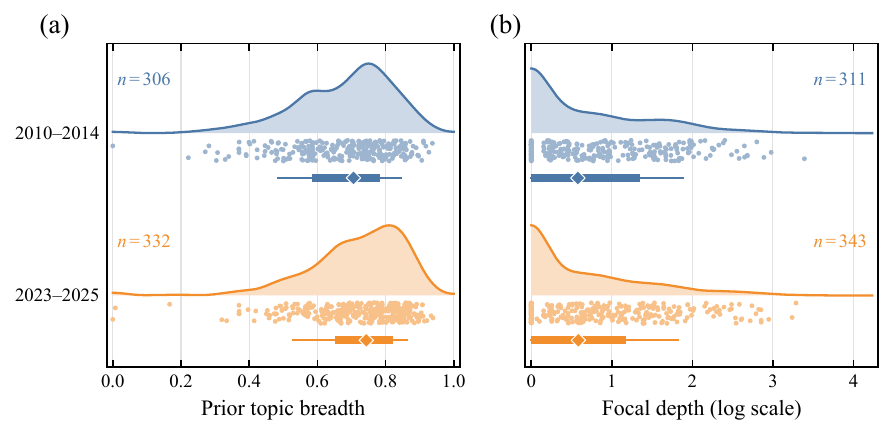}
\caption{Earlier and recent distributions of individual expertise. Blue shows 2010--2014 and orange shows 2023--2025. (a) Prior topic breadth, with 306 and 332 observed contributor events. (b) Focal depth on its logarithmic scale, with 311 and 343 events. The filled shapes show weighted kernel densities, small points show every observed event, diamonds mark weighted medians, thick segments span the interquartile ranges, and thin segments span the 10th--90th percentiles. Each field receives equal total weight within each period, with sampling weights retained within fields. Both periods use the same smoothing bandwidth within each panel.}
\label{fig:period_distributions}
\end{figure*}

The distributions in Figure~\ref{fig:period_distributions} show how these period contrasts relate to the spread of observed individual profiles. After balancing the field contributions, median prior breadth increases from 0.706 to 0.744, while its lower quartile rises from 0.585 to 0.653. The shift is therefore visible through the central part of the breadth distribution. Focal depth has nearly unchanged medians of 0.578 and 0.585, while its upper quartile moves from 1.350 to 1.178. The substantial overlap in depth and the moderate displacement in breadth place recent broadening within a continuing diversity of individual research profiles. These descriptive distributions complement the covariate adjusted contrasts in Figure~\ref{fig:individual}.

\subsection{Focal depth was associated with higher mature citation impact}

\begin{figure*}[!tbp]
\centering
\includegraphics[width=0.92\textwidth]{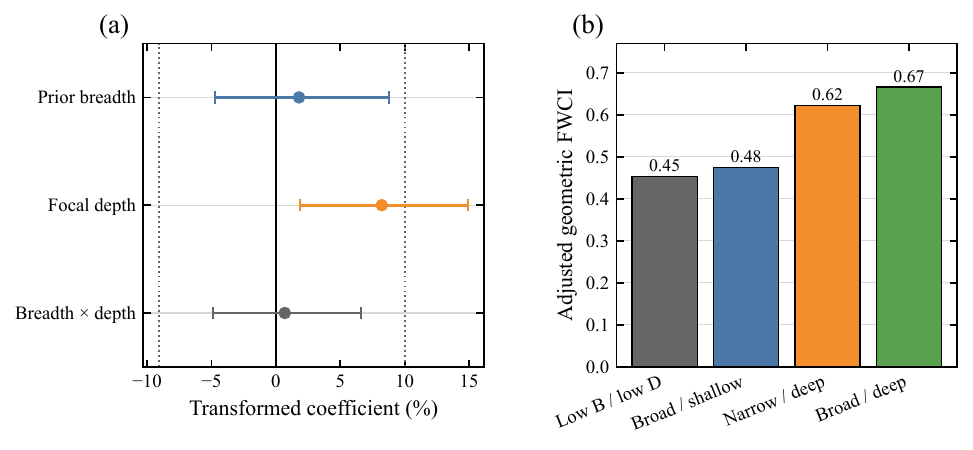}
\caption{Conditional citation returns to prior capability. (a) Transformed coefficients $100[\exp(\hat\beta)-1]$ for breadth, focal depth, and their interaction, with HC3 95\% confidence intervals. Breadth and depth are standardized within field; each main coefficient is evaluated at the other variable's mean. The interaction is the coefficient of their product. Dotted lines mark the sensitivity bounds $\pm\log(1.10)$, transformed to --9.1\% and 10.0\%. (b) Fitted geometric FWCI values at the observed lower and upper quartiles of standardized breadth and depth, obtained by averaging fitted log outcomes over the observed covariate distribution and applying $\exp(\cdot)-1$. Bars summarize fitted profiles.}

\label{fig:returns}
\end{figure*}

The mature return model connects prior individual profiles to the subsequent recognition of the focal contribution. Among 959 complete contributor events published through 2022, a one standard deviation increase in focal depth was associated with 8.2\% higher $1+\mathrm{FWCI}$ [1.9\%, 14.9\%] ($p=0.010$), conditional on mean breadth within the field. The model adjusts for field and year, career age, recent publication volume, team size, and open access status. As Figure~\ref{fig:returns}a illustrates, the estimated breadth association at mean depth was 1.8\% [--4.7\%, 8.8\%] ($p=0.598$), and the interaction was 0.7\% [--4.9\%, 6.6\%] ($p=0.814$, Holm $p=1.000$). The positive depth coefficient links accumulated knowledge of the focal subfield to scientific recognition after accounting for the amount of recent output and collaborative scale.

Equivalence checks sharpen the interpretation of the smaller coefficients. Under the sensitivity bounds of $\pm\log(1.10)$, breadth ($p_{\mathrm{TOST}}=0.011$) and the interaction ($p_{\mathrm{TOST}}=0.001$) fall within the specified range of practically small average associations. On the percentage scale in Figure~\ref{fig:returns}a, these bounds are --9.1\% and 10.0\%. The interaction describes how the slope associated with one capability changes as the other increases by one standard deviation. Its estimate supports a return surface with limited average complementarity over the observed profiles.

The profile comparison in Figure~\ref{fig:returns}b translates the fitted surface into interpretable combinations. Evaluated at the lower and upper quartiles of breadth and depth, adjusted geometric FWCI was 0.45 for low breadth and low depth, 0.48 for broad and shallow, 0.62 for narrow and deep, and 0.67 for broad and deep profiles. Raising depth at either breadth level corresponds to a larger fitted separation than raising breadth at either depth level. These standardized fitted values summarize the shape of the same fitted model, with all remaining covariates held at their observed values. The broad and deep profile attains the highest fitted value, while the narrow and deep profile already captures much of the fitted advantage.

The depth association remained positive under ten year histories, the last ten works, ORCID restriction, a strict definition of focal depth, unweighted estimation, citation percentile, and a bootstrap stratified by field and year. The bootstrap 95\% interval was [0.0206, 0.1344] on the log scale. Estimates weakened for raw citations through 2019 and for a top decile outcome, indicating that recognition measured over the continuous normalized citation distribution is the clearest expression of the association. These checks support focal accumulation as a persistent correlate of impact across several ways of measuring a contributor's prior expertise.

\begin{table*}[!t]
\centering
\caption{Six primary tests with familywise adjustment. The family was specified after data construction and before estimating capability and outcome associations.}
\label{tab:frozen}
\begin{threeparttable}
\begin{tabularx}{\textwidth}{@{}Yrrrrl@{}}
\toprule
Test & Estimate & 95\% CI & Raw $p$ & Holm $p$ & $H_0$ at 5\% \\
\midrule
Annual paper topic breadth trend & -0.0012 & [-0.0017, -0.0007] & $<0.001$ & $<0.001$ & Reject \\
Annual $\log(1+\mathrm{team\ size})$ trend & 0.0174 & [0.0160, 0.0188] & $<0.001$ & $<0.001$ & Reject \\
Individual breadth, 2019--2022 vs. 2010--2014 & 0.0190 & [-0.0078, 0.0459] & 0.165 & 0.494 & Do not reject \\
Individual depth, 2019--2022 vs. 2010--2014 & -0.0274 & [-0.1451, 0.0902] & 0.648 & 1.000 & Do not reject \\
Breadth $\times$ depth return on log FWCI & 0.0069 & [-0.0504, 0.0641] & 0.814 & 1.000 & Do not reject \\
Mean post-2022 forecast residual & -0.0108 & [-0.026, 0.004] & 0.116 & 0.463 & Do not reject \\
\bottomrule
\end{tabularx}
\begin{tablenotes}[flushleft]\footnotesize
\item HC3 intervals are reported except for the post-2022 residual, whose interval and test use the six field level mean residuals. Table~\ref{tab:effects} supplements these decisions with effect sizes and equivalence analyses.
\end{tablenotes}
\end{threeparttable}
\end{table*}

\begin{table*}[!t]
\centering
\caption{Effect sizes and equivalence analyses.}
\label{tab:effects}
\begin{threeparttable}
\begin{tabular}{@{}lrrl@{}}
\toprule
Quantity & Estimate & 95\% CI & Scale \\
\midrule
Paper topic breadth, cumulative 2010--2022 & -0.0144 & [-0.0199, -0.0089] & distance \\ 
Cited knowledge breadth, cumulative 2010--2022 & 0.0273 & [0.0002, 0.0545] & distance \\ 
Team size, cumulative 2010--2022 (auxiliary model) & 37.3 & [34.4, 40.3] & \% \\ 
Individual breadth, 2019--2022 vs. baseline & 0.0190 & [-0.0078, 0.0459] & distance \\ 
Individual breadth, 2023--2025 vs. baseline & 0.0327 & [0.0044, 0.0610] & distance \\ 
Return to breadth per within field SD & 1.8 & [-4.7, 8.8] & \% \\ 
Return to depth per within field SD & 8.2 & [1.9, 14.9] & \% \\ 
Return to breadth $\times$ depth & 0.7 & [-4.9, 6.6] & \% \\ 
\bottomrule
\end{tabular}
\vspace{0.6em}
\begin{tabular}{@{}lrrr@{}}
\toprule
Equivalence target & SESOI & 90\% CI & TOST $p$ \\
\midrule
Individual breadth change by 2019--2022 & $\pm$0.050 & [-0.004, 0.042] & 0.012 \\
FWCI breadth return & $\pm$0.095 & [-0.038, 0.073] & 0.011 \\
FWCI breadth $\times$ depth interaction & $\pm$0.095 & [-0.041, 0.055] & 0.001 \\
\bottomrule
\end{tabular}
\begin{tablenotes}[flushleft]\footnotesize
\item The team size percentage is $100[\exp(12\hat\beta)-1]$ from the auxiliary $\log(\text{team size})$ model among 38,285 papers with at least one recorded author; the specified test instead models $\log(1+\text{team size})$. Percentage returns are $100[\exp(\hat\beta)-1]$ for $1+\mathrm{FWCI}$. The cited knowledge HC3 interval is shown in the first panel; clustering within field and year cells gives an annual slope of 0.00228 [--0.00053, 0.00509], $p=0.112$. SESOI denotes the smallest effect size of interest, in distance units for individual breadth and log units of $1+\mathrm{FWCI}$ for citation returns. These bounds were set during robustness analysis, and the two one sided tests (TOST) are sensitivity analyses. The main breadth and depth coefficients are evaluated at the other capability's within field mean; the interaction coefficient multiplies the product of the two standardized variables.
\end{tablenotes}
\end{threeparttable}
\end{table*}

Table~\ref{tab:frozen} brings the six primary tests together. The trends in team size and paper topic breadth retain statistical support after Holm adjustment, locating the clearest systematic changes at the organizational and output levels. Table~\ref{tab:effects} complements these tests with cumulative changes and equivalence analyses. Reading significance alongside magnitude makes the substantive pattern clearer. Collaboration expands appreciably, output topics concentrate, and individual breadth and its average citation association change on more modest scales under the stated sensitivity bounds.

\subsection{Recent patterns in the context of longer trajectories}

\begin{figure*}[!tbp]
\centering
\includegraphics[width=0.944116\textwidth]{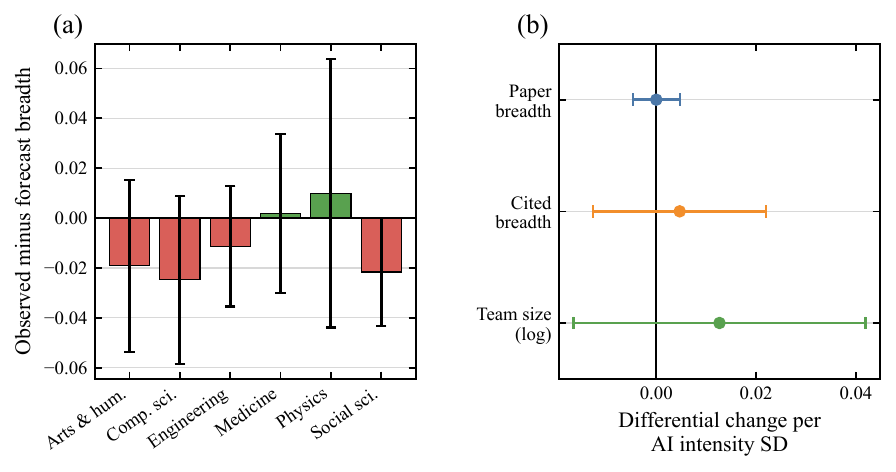}
\caption{Recent patterns relative to earlier trajectories and baseline AI intensity. (a) Mean 2023--2025 paper breadth minus the forecast from each field's selected 2010--2018 trend model. Bars show 1.96 standard errors of the residuals across the three recent years; the pooled test in the text uses the six field means. (b) Continuous difference in differences coefficients per standard deviation of pre-2019 AI intensity within field, with 95\% confidence intervals clustered by subfield. The outcomes are paper topic breadth, cited knowledge breadth, and $\log(1+\text{team size})$. Coefficients retain each outcome's measurement scale.}

\label{fig:counterfactual}
\end{figure*}

Figure~\ref{fig:counterfactual}a compares recent paper breadth with forecasts selected using earlier publication years. For each field, linear and quadratic models were fitted to 2010--2018 data and selected by their 2019--2022 forecast error before predicting 2023--2025. Four fields fell below their projected paths and two rose above them. The mean residual was --0.01084 [--0.026, 0.004] ($p=0.116$, Holm $p=0.463$), while placebo origins in 2017, 2019, and 2020 produced mean residuals near zero. The mixture of field directions places recent developments within heterogeneous trajectories and favors interpreting the pooled contraction as a longer evolution in research organization.

The complementary comparison in Figure~\ref{fig:counterfactual}b asks whether recent changes were larger in subfields historically closer to AI. Baseline intensity is the smoothed logit of each subfield's 2010--2018 share of papers with an AI topic, standardized within broad field. Across 83 subfields, continuous difference in differences with subfield and year fixed effects yielded a recent period coefficient of 0.00000 for paper topic breadth [--0.00467, 0.00467] ($p=1.000$), 0.0127 for $\log(1+\text{team size})$ [--0.0166, 0.0419] ($p=0.396$), and 0.0047 for cited knowledge breadth [--0.0127, 0.0220] ($p=0.596$). The tightly centered topic estimate is consistent with similar recent changes across the baseline intensity gradient, while the other two outcomes retain wider ranges of plausible differential change. A legacy core AI vocabulary produced closely matching estimates, as shown in Table~\ref{tab:s_ai_definitions}.

Together, the forecast and intensity comparisons locate the observed reorganization across both calendar time and subfield context. The former evaluates departures from each field's earlier trajectory; the latter evaluates heterogeneity across prior AI intensity. For individual breadth, the exploratory intensity contrast is positive at 0.0235 [--0.0005, 0.0476], as reported in Table~\ref{tab:s_ai}. This pattern identifies individual repertoires as a useful level for subsequent monitoring, alongside the more firmly established expansion of teams and concentration of output topics.

\section{Discussion}

The evidence supports a differentiated view of scientific expertise. Across six fields, research became more collaborative and its output topics more concentrated, while cited knowledge remained stable to modestly broader. Established contributors' repertoires changed gradually, and focal depth retained a positive association with mature citation impact. These findings position breadth and depth at complementary levels of the research process. A scientist can contribute through focused accumulation while drawing on knowledge and collaborators whose reach extends beyond that specialty.

The divergence between paper topics and cited inputs is central to this interpretation. Figure~\ref{fig:trends} shows that the intellectual span of a contribution and the distance among its knowledge sources can evolve differently. A specialized question may require methods, evidence, or conceptual resources drawn from several domains. The resulting article can remain focused even as its route to that result traverses a broader literature. This pattern connects earlier observations of local specialization and wider integration \parencite{porter2009,evans2008} to a contemporary debate about AI and scientific work. Search tools and AI provide plausible channels for extending access to inputs while scientific problem formulation sustains the coherence of outputs.

Team expansion supplies the organizational counterpart to this separation. The modelled 37.3\% growth in team size, together with the persistence of topic concentration after standardizing team composition, is consistent with more contributors coordinating around focused problems. In the burden of knowledge account, collaboration enables researchers to combine specialized contributions as the frontier becomes harder to master individually \parencite{jones2009,wuchty2007}. Our multilevel evidence extends that account by showing that growing collaborative scale can coexist with modest changes in the measured breadth of established contributors. This makes the distribution of expertise across research participants a promising explanation for how contemporary science manages expanding knowledge.

The citation models add an evaluative dimension. Focal accumulation is associated with higher mature impact across several history and identity definitions, and the fitted profiles show substantial separation between shallow and deep contributors at both breadth levels. This finding aligns with evidence of persistent specialization rewards \parencite{rassenfosse2022} and the importance of domain expertise in changing knowledge environments \parencite{teodoridis2019}. Breadth may contribute through exploration, unusual combinations, and translation across contexts, whose rewards can be heterogeneous \parencite{uzzi2013,foster2015,leahey2017}. The present estimates place the clearest average citation association in focal depth and identify the conditions under which breadth strengthens that association as a valuable next question.

The temporal evidence further places AI within an ongoing organizational evolution. Both focused output and expanding collaboration were established before widespread generative AI access. The recent forecasts and baseline intensity contrasts place the findings within heterogeneous field trajectories over the observed horizon. Studies of AI related research and textually inferred LLM use locate productivity and impact associations within that broader setting \parencite{hao2026,bianchini2026,kusumegi2025}. These levels of evidence are complementary because changes in how particular researchers perform tasks can precede shifts in the aggregate structure of publications and careers.

For scientific training and evaluation, the findings support deep expertise as an empirical anchor, complemented by the capacity to locate, evaluate, and connect knowledge beyond the focal specialty. Search, rapid learning, and coordination are candidate capabilities through which researchers can use broader inputs while sustaining reliable specialist contributions. This interpretation develops the case for recalibrating expertise in the presence of AI \parencite{lin2026} and for preserving the scientific judgment needed to assess model outputs \parencite{messeri2024}. Public contribution statements, method histories, and records of research software offer routes to testing these capabilities more directly in subsequent computational work.

The methodological implication is equally substantive. Different measures of breadth answer different questions about the organization of science, as emphasized by studies of interdisciplinarity measurement \parencite{calatrava2016,zwanenburg2022,cantone2024}. Linking output topics, cited inputs, contributor histories, and collaborative scale makes their disagreement interpretable. The resulting picture is one of focused expertise embedded in an increasingly collaborative research system, with breadth expressed partly through access to knowledge beyond the individual publication repertoire.

\section{Methods}

\subsection{Data source and sampling by field and year}

OpenAlex is a fully open scholarly knowledge graph connecting works, authors, institutions, topics, and citation links \parencite{priem2022}. The analysis uses records retrieved on 12 September 2026, with publication years restricted to 2010--2025. OpenAlex is well suited to transparent research on the science of science, although its author disambiguation and automated topics remain evolving data products. Related open infrastructures such as SciSciNet demonstrate the value of versioned, reusable scholarly graphs \parencite{sciscinet2023}.

We selected six broad OpenAlex fields spanning different epistemic and authorship regimes, comprising arts and humanities, computer science, engineering, medicine, physics and astronomy, and social sciences. For each field and publication year from 2010 through 2025, we drew 500 article records with a fixed random seed for each field and year, yielding 48,000 works. We retained title, date, language, authorships, topics, primary topic, references, citation metrics, and open access status. Removing retracted and paratext records left 47,959 articles. Every sampled field agreed with the article's primary field. The design gives each combination of field and year equal representation, so the reported coefficients describe a balanced trend across the six fields.

\subsection{Topic hierarchy and paper breadth}

OpenAlex assigns up to three scored topics inferred by a model to a work; each topic belongs to a subfield and field \parencite{openalextopics2026}. For topics $i$ and $j$, we define the hierarchical distance
\begin{equation}
d(i,j)=
\begin{cases}
0, & i=j,\\
1/3, & \text{same subfield},\\
2/3, & \text{same field but different subfields},\\
1, & \text{different fields}.
\end{cases}
\end{equation}
If $p_{wi}$ is topic $i$'s score divided by the sum of a paper's topic scores, paper breadth is
\begin{equation}
B^{P}_{w}=\sum_i\sum_j p_{wi}p_{wj}d(i,j).
\end{equation}
This is a Rao--Stirling integration of variety, balance, and disparity \parencite{stirling2007}. The primary trend sample contains exactly three topics per paper, preventing the changing share of records with one, two, or three topics from mechanically determining the result. Robustness analyses use equal topic weights, maximum pair distance, cross field indicators, and all works with topic count fixed effects.

\subsection{Independent cited knowledge validation}

Reference completeness varied substantially in the full sample, especially in arts and humanities. We therefore used a separate validation balanced across fields and years. The initial pilot using fixed random ranks sampled 10 papers with references in each field and year. After observing that only 623 pre-2023 papers met the coverage rule and that all three trend intervals were too wide to distinguish small changes from zero, we specified a precision expansion comprising 35 papers with at least five listed references in every combination of field and year and a maximum of 20 references per paper selected by fixed random ranks. Selection of articles and their references used identifiers alone.

The expansion contained 3,360 papers and 55,223 unique identifiers of cited works. We resolved 51,736 IDs (93.7\%) through the OpenAlex API. A focal paper was eligible if at least five sampled references, accounting for at least 70\% of its sampled identifiers, resolved to works with a classified primary topic, leaving 3,215 papers overall and 2,619 through 2022. With $m_w$ resolved references and $z_r$ the primary topic of reference $r$, cited knowledge breadth is the unbiased mean across distinct reference pairs
\begin{equation}
B^{R}_{w}=\frac{2}{m_w(m_w-1)}\sum_{r<s}d(z_r,z_s).
\end{equation}
We also calculated the effective number of cited fields, the share outside the focal paper's primary field, and a coarser field/subfield distance. Figure~\ref{fig:scoverage} reports coverage by field and year, allowing the reference trends to be assessed alongside changes in metadata availability.

\subsection{Contributor sampling and prior histories}

Sampling authorship records gives greater representation to large teams, while selecting the first or last author depends on disciplinary conventions. We instead selected a focal article by fixed random rank and then selected one contributor uniformly from teams of 1--20. The initial frame contained 100 events per field in four periods (2010--2014, 2015--2018, 2019--2022, and 2023--2025), or 2,400 events. We resolved author metadata and restricted the history collection frame to authors with 3--300 lifetime works, a rule chosen to avoid both sparse histories and unusually prolific or potentially conflated identities. Taking all eligible reserve events yielded 1,755 events and 1,754 unique author IDs.

For each selected author, we downloaded complete OpenAlex histories from 2000 through 2025 for articles, preprints, and proceedings articles. Every capability at focal year $t$ uses only works with publication year strictly less than $t$. Main eligibility requires at least three prior works spanning at least two years. We flag an author identity as unstable when more than 30 prior works occur in one year; 1,316 events satisfy both the history and identity rules. Weighting accounts for the probability of article selection within each field and period, the probability of inclusion in the history frame, and the uniform draw of an author within each team. The main estimand is consequently a typical sampled paper's randomly chosen established contributor, conditional on the stated history frame.

\subsection{Individual breadth and depth}

Within the five years before a focal paper, prior work $k$ receives recency weight $\exp[-\lambda(t-y_k)]$, where $\lambda=\log(2)/5$ and $y_k$ is its publication year. This is multiplied by the normalized score of each topic on $k$. Aggregating and normalizing those masses produces a topic distribution for each author $q_{ait}$. Individual breadth is
\begin{equation}
B^{I}_{at}=\sum_i\sum_j q_{ait}q_{ajt}d(i,j).
\end{equation}
Let $s(w)$ denote the primary subfield of focal paper $w$. Focal depth preserves the unnormalized topic mass in that subfield after weighting for recency,
\begin{equation}
D_{aw}=\log\left[1+\sum_{k<t}\exp\{-\lambda(t-y_k)\}
\sum_{i:s(i)=s(w)}p_{ki}\right].
\end{equation}
This definition allows breadth and depth to vary independently, so that an author can be simultaneously broad and deep. Robustness definitions use ten year histories, the last ten prior works, and only primary subfield matches.

The direct visual comparisons use the same 2010--2014 baseline and 2023--2025 recent period. Field endpoints average annual field means with equal weights for years. For individual distributions, we retain established contributors with stable identities and an observed value of the plotted metric. We normalize the original contributor sampling weights separately within each field and period so that each of the six fields contributes equal total weight. Medians and percentile ranges are obtained from the weighted empirical distribution. Gaussian kernel densities use a common bandwidth across periods for each metric, equal to the pooled weighted standard deviation multiplied by the effective sample size to the power $-1/5$. Reflection at zero, and at one for breadth, respects the support boundaries. The plots retain every observed event; vertical jitter separates overlapping points. These displays compare repeated cross sections, whereas the period regressions additionally adjust for career age and recent publication volume.

\subsection{Trend and return models}

Paper models regress breadth or $\log(1+\text{team size})$ on a linear year term and field fixed effects through 2022. Main standard errors are HC3; one way clustering within the 78 field and year cells provides a robustness check. For a proportional effect on team size itself, an auxiliary model replaces $\log(1+\text{team size})$ with $\log(\text{team size})$ and omits records with no OpenAlex authorship. The team composition decomposition reweights recent exactly three topic papers within each field to the field's 2010--2014 team size category shares.

Contributor trend models are weighted least squares with period indicators, field fixed effects, a spline with four degrees of freedom in career age, and $\log(1+$number of recent prior works$)$. HC3 standard errors allow for unequal residual variance. The two primary contrasts compare 2019--2022 with 2010--2014. Estimates for recent years and alternative histories are sensitivity analyses.

OpenAlex FWCI counts citations in the publication year and following three years and normalizes by document type, publication year, and subfield \parencite{openalexfwci2026}. To give every focal work that full opportunity window, return models end in 2022. FWCI is winsorized at the 99th percentile and transformed as $\log(1+\mathrm{FWCI})$. Within each broad field, $B^I$ and $D$ are standardized to $B_z$ and $D_z$. The model takes the form
\begin{equation}
\begin{aligned}
\log(1+\mathrm{FWCI}_{w})={}&\beta_B B_z+\beta_D D_z+\beta_{BD}B_zD_z\\
&+X_w\gamma+\alpha_{f\times t}+\varepsilon_w,
\end{aligned}
\end{equation}
where $X$ contains career age, recent output, team size, and open access status, and $\alpha_{f\times t}$ is a fixed effect for each combination of field and year. We use the contributor sampling weight and HC3 errors. Robustness checks replace the outcome with top decile status, citation percentile, or log citations; alter the history window; restrict to authors linked to ORCID; omit weights; and leave out each field. A 400-draw bootstrap resamples observations within cells defined by field and year.

\subsection{Post-2022 counterfactuals and AI intensity}

The predictive counterfactual separates model selection from evaluation. For each field, linear and quadratic trends were fitted to 2010--2018 annual paper topic breadth. The form with lower root mean squared error in 2019--2022 was selected and extrapolated to 2023--2025. The primary statistic is the mean recent period residual within each field, tested across the six fields. Placebo cutoffs in 2017, 2019, and 2020 diagnose whether the procedure produces systematic false breaks. The resulting contrast measures departures from earlier trajectories over a period of widespread generative AI access.

For exploratory heterogeneity, a paper was marked AI related when any current OpenAlex topic name contained one of ten strings covering artificial intelligence, machine learning, deep learning, neural networks, natural language processing, computer vision, generative adversarial networks, large language models, foundation models, and transformer models. For each primary subfield, we calculated its 2010--2018 share, applied Jeffreys smoothing, took the logit, and standardized within broad field. We retained 83 subfields with at least 30 baseline and 15 post-2022 sampled papers. Continuous difference in differences models for 2019--2025 include subfield and year fixed effects, an interaction between standardized baseline AI intensity and the post-2022 indicator, and standard errors clustered by subfield. Event study interactions use 2022 as the reference year and test 2019--2021 coefficients jointly. The exposure interaction summarizes a conditional gradient in recent change. Its counterfactual interpretation requires parallel trends and assumptions linking continuous exposure contrasts to treatment responses \parencite{callaway2024,rambachan2023}. Joint pretrend tests gave $p=0.750$ for paper breadth, $p=0.309$ for team size, and $p=0.074$ for reference breadth. For the latter, event study and period average estimates differed, indicating sensitivity to the temporal specification.

\subsection{Multiplicity and equivalence}

The six primary $p$ values were adjusted together with Holm's method. The hypothesis family was specified after data construction and before estimating the capability and outcome models. Reference validation used an adaptive precision expansion following inspection of the pilot's uncertainty, with an independent holdout excluding pilot articles. Team selection, aggregated trends, and AI vocabulary comparisons were added during manuscript review as supplementary sensitivity analyses.

Robustness analyses apply two one sided tests to assess practical magnitude. The smallest effects of interest are $\pm0.05$ on the individual 0--1 breadth scale and $\pm\log(1.10)$ for the standardized breadth and interaction coefficients in the log FWCI model. These bounds were chosen during robustness analysis after specifying the six primary tests. They therefore serve as sensitivity thresholds. Equivalence is assessed using 90\% intervals, corresponding to two one sided tests at the 5\% level. The log scale bounds transform to percentage changes of --9.1\% and 10.0\% in $1+\mathrm{FWCI}$.

\subsection{Scope of measurement and inference}

The estimates characterize deployed knowledge and citation recognition within a balanced six field sample. Contributor analyses target a sampled article's established contributor, conditional on teams of 1--20 recorded authors and the history frame of 3--300 lifetime works. The individual estimand therefore covers established contributors in this team size range; hyperauthorship teams remain represented in the article analyses. This sampling structure gives each field and year equal representation and emphasizes active contributors with observable prior research. Citation coefficients are conditional associations; they may combine the contribution of expertise with topic selection, resources, and cumulative advantage. The forecast and AI intensity designs respectively describe temporal departures and heterogeneity by historical proximity to AI research. Identifying effects of actual generative AI use would require observations of adoption and a credible comparison trajectory.

All topics are measured using the same retrieval snapshot, including retrospective classifications. Topic counts, semantic assignment, reference coverage, and author disambiguation therefore shape the observed measures. Exactly three topic analyses, alternative distances, independent reference sampling, ORCID restrictions, and alternative history windows assess these sources of sensitivity. The six fields span diverse publication practices, with coverage strongest for indexed articles and more variable for book based scholarship. The recent period covers three publication years, while citation returns use complete windows through 2022. These choices define the population and horizon over which the multilevel findings apply.

\section*{Data and code availability}

The study uses public OpenAlex records. Analysis code and data required to reproduce the reported results are organized in a replication package for deposit in a public repository before journal submission.

\section*{AI assistance disclosure}

Generative AI assisted literature discovery, code implementation, diagnostic design, figure preparation, and language editing under human direction. The human authors conceived the research questions and core ideas, established the analytical framework, selected the data and methods, directed and reviewed all code development, evaluated robustness, interpreted the results, designed the figures and tables, structured the manuscript logic, formulated the conclusions, verified the sources and references, and determined the final scientific expression. The human authors take full responsibility for the research and the submitted manuscript.

\section*{Acknowledgments}

No external funding or conflicts of interest are declared.

\FloatBarrier
\appendix
\titleformat{\section}[hang]{\normalfont\Large\bfseries\filright}{\appendixname~\thesection}{0.8em}{}
\counterwithin{figure}{section}
\counterwithin{table}{section}
\renewcommand{\thefigure}{\thesection\arabic{figure}}
\renewcommand{\thetable}{\thesection\arabic{table}}
\section{Supplementary analyses}

\subsection{Field variation in measurement divergence}

\begin{figure*}[!tbp]
\centering
\includegraphics[width=\textwidth]{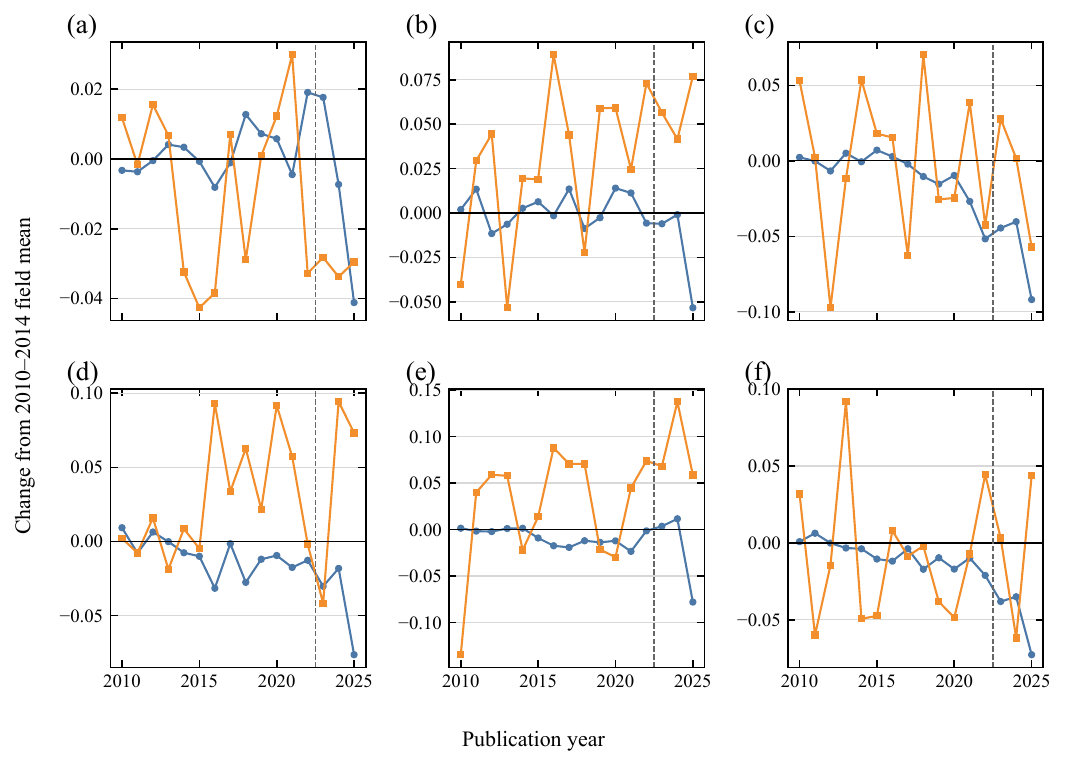}
\caption{Topic and cited knowledge breadth within fields. Blue solid lines with circular markers show paper topic breadth, and orange solid lines with square markers show cited knowledge breadth. Each series is centered on its own 2010--2014 field mean. Black horizontal lines indicate zero change from this baseline, and gray vertical dashed lines mark the start of 2023. Panels correspond to (a) arts and humanities, (b) computer science, (c) engineering, (d) medicine, (e) physics and astronomy, and (f) social sciences.}

\label{fig:sfields}
\end{figure*}

Figure~\ref{fig:sfields} decomposes the pooled breadth trajectories into the six constituent fields. Engineering, medicine, physics and astronomy, and social sciences have negative paper topic slopes; computer science is approximately flat, and arts and humanities slightly positive. The cited knowledge paths exhibit more annual variation, consistent with the smaller reference sample in each cell. Reading the panels jointly shows that output concentration recurs across several fields, while the input trajectory remains sensitive to disciplinary context. Table~\ref{tab:s_topic} reinforces the output result across alternative topic distances, weighting rules, and treatments of topic count, connecting the pooled finding to multiple definitions of topical span.

\begin{table*}[!b]
\centering
\caption{Robustness of the 2010--2022 paper topic breadth trend.}
\label{tab:s_topic}
\begin{tabular}{@{}lrrrr@{}}
\toprule
Specification & Annual slope & 95\% CI & $p$ & $N$ \\
\midrule
Primary score weighted, exactly 3 topics & -0.00120 & [-0.00166, -0.00074] & $<0.001$ & 29,992 \\
Exactly 3 topics, SE clustered within field and year cells & -0.00120 & [-0.00190, -0.00050] & $<0.001$ & 29,992 \\
All papers, topic count adjusted & -0.00099 & [-0.00136, -0.00061] & $<0.001$ & 38,976 \\
Maximum topic distance, exactly 3 topics & -0.00098 & [-0.00161, -0.00036] & 0.002 & 29,992 \\
Unweighted pair distance, exactly 3 topics & -0.00117 & [-0.00173, -0.00061] & $<0.001$ & 29,992 \\
Any cross-field topic (linear probability) & -0.00268 & [-0.00409, -0.00127] & $<0.001$ & 29,992 \\
\bottomrule
\end{tabular}
\end{table*}

\begin{table*}[!t]
\centering
\caption{Robustness of capability and citation estimates. Entries are log-coefficients with two sided $p$-values in parentheses.}
\label{tab:s_returns}
\begin{tabular}{@{}lrrrr@{}}
\toprule
Specification & Breadth & Depth & Breadth $\times$ depth & $N$ \\
\midrule
Five year main & 0.018 (0.598) & 0.079 (0.010) & 0.007 (0.814) & 959 \\
Ten year history & 0.011 (0.729) & 0.072 (0.015) & 0.007 (0.790) & 970 \\
Last ten works & -0.003 (0.942) & 0.074 (0.035) & 0.018 (0.599) & 685 \\
ORCID linked authors & 0.049 (0.270) & 0.136 (0.002) & 0.026 (0.521) & 601 \\
Unweighted & 0.007 (0.811) & 0.108 ($<0.001$) & -0.011 (0.653) & 959 \\
Strict primary subfield depth & 0.018 (0.583) & 0.093 (0.002) & 0.003 (0.924) & 959 \\
\bottomrule
\end{tabular}
\end{table*}

\begin{table*}[!t]
\centering
\caption{Exploratory post-2022 heterogeneity by pre-2019 AI intensity.}
\label{tab:s_ai}
\begin{tabular}{@{}lrrrrr@{}}
\toprule
Outcome & Post $\times$ AI intensity & 95\% CI & $p$ & $N$ & Subfields \\
\midrule
Paper topic breadth & 0.0000 & [-0.0047, 0.0047] & 1.000 & 16,591 & 83 \\
$\log(1+\mathrm{team\ size})$ & 0.0127 & [-0.0166, 0.0419] & 0.396 & 20,393 & 83 \\
Cited knowledge breadth & 0.0047 & [-0.0127, 0.0220] & 0.596 & 1,360 & 83 \\
Individual prior breadth & 0.0235 & [-0.0005, 0.0476] & 0.055 & 661 & 77 \\
Individual focal depth & -0.0467 & [-0.1719, 0.0785] & 0.465 & 674 & 77 \\
\bottomrule
\end{tabular}
\end{table*}

\subsection{Reference validation and coverage}

The initial pilot yielded an annual reference breadth slope of --0.00048 [--0.00517, 0.00422] ($p=0.842$; $N=623$). The precision expansion gave 0.00228 [0.00002, 0.00454] under HC3 errors and [--0.00053, 0.00509] with clustering within field and year cells. Excluding all 187 articles overlapping the pilot left 2,469 eligible pre-2023 articles and a slope of 0.00255 [0.00024, 0.00486] under HC3 errors. The holdout reproduces the positive expanded estimate using separate focal articles, while the clustered interval supports the synthesis of stable to modestly broader cited knowledge.

\begin{figure}[!tbp]
\centering
\includegraphics[width=0.98\columnwidth]{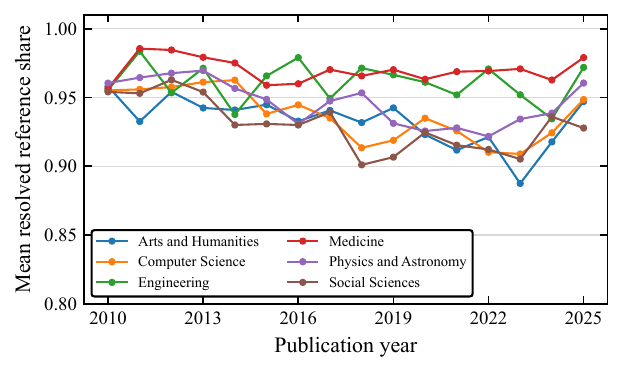}
\caption{Reference metadata coverage by field and publication year. Lines show the mean share of sampled reference identifiers resolved to a work with a primary topic. Coverage is calculated for each focal article before averaging within field and year.}
\label{fig:scoverage}
\end{figure}

Figure~\ref{fig:scoverage} displays the field and year structure of reference metadata coverage. The aggregate resolution rate of 93.7\% describes unique cited identifiers, whereas the figure summarizes each article's share of sampled references with usable primary topics. These quantities answer different coverage questions. The eligibility rules require at least five classified references and 70\% coverage by classified references, preserving a minimum information base for pairwise distance estimates. Field trajectories provide context for assessing whether differences in metadata availability accompany the observed reference breadth patterns.

\subsection{Stability of the capability and citation association}

Table~\ref{tab:s_returns} compares the fitted coefficients across history length, publication count, identity quality, sampling weights, and the definition of depth. The positive depth association persists across these choices, indicating that it reflects a recurring relationship between prior accumulation and impact. Bootstrap 95\% percentile intervals were [--0.0500, 0.0820] for breadth, [0.0206, 0.1344] for depth, and [--0.0421, 0.0574] for the interaction on the log scale. Omitting each field in turn preserved a positive depth coefficient, ranging from 0.0575 to 0.1284, showing that the association is distributed across the sampled research settings.

The interaction remained centered near zero across the field omission models. One exploratory dynamic specification estimated a nominal negative interaction change in 2019--2022. Considered alongside the primary interaction and the broader sensitivity results, the overall pattern remains one of limited average complementarity with a positive focal depth association. This combination motivates more targeted study of the research contexts in which breadth becomes especially productive.

\subsection{Recent heterogeneity by prior AI intensity}

Table~\ref{tab:s_ai} extends the intensity analysis to individual capabilities. Prior breadth has a positive estimate of 0.0235 per standard deviation of baseline AI intensity [--0.0005, 0.0476], whereas focal depth has a negative estimate of --0.0467 [--0.1719, 0.0785]. The positive breadth estimate motivates closer examination of contributor repertoires in subfields with greater prior AI intensity. Its interval spans both little differential change and more appreciable broadening, providing a quantitative range for future observations. These exploratory contrasts describe heterogeneity by historical AI proximity within the model and sample defined in Methods.

\begin{table*}[!t]
\centering\small
\begin{threeparttable}
\caption{Team coverage and sensitivity of the paper topic trend.}
\label{tab:s_selection}
\begin{tabular}{@{}lrr@{}}
\toprule
Field & 2010--2014, $n/N$ (\%) & 2023--2025, $n/N$ (\%) \\
\midrule
Arts and Humanities & 2/2,500 (0.08) & 0/1,499 (0.00) \\
Computer Science & 2/2,497 (0.08) & 2/1,488 (0.13) \\
Engineering & 2/2,500 (0.08) & 0/1,500 (0.00) \\
Medicine & 10/2,500 (0.40) & 27/1,497 (1.80) \\
Physics and Astronomy & 44/2,500 (1.76) & 57/1,499 (3.80) \\
Social Sciences & 1/2,500 (0.04) & 1/1,500 (0.07) \\
All six fields & 61/14,997 (0.41) & 87/8,983 (0.97) \\
\bottomrule\end{tabular}
\medskip
\begin{tabular}{@{}lrrrr@{}}
\toprule
Trend specification & Annual slope & 95\% CI & $p$ & Units \\
\midrule
All exactly three topic papers & $-0.00120$ & $[-0.00166, -0.00074]$ & $<0.001$ & 29,992 \\
Teams of 1--20 authors & $-0.00122$ & $[-0.00169, -0.00076]$ & $<0.001$ & 29,329 \\
Equal field and year cell means & $-0.00111$ & $[-0.00184, -0.00038]$ & 0.003 & 78 \\
Mean of six field slopes & $-0.00111$ & $[-0.00266, 0.00044]$ & 0.125 & 6 \\
\bottomrule\end{tabular}
\begin{tablenotes}\footnotesize\item The upper panel reports papers with more than 20 recorded authors, with all eligible articles as the denominator. The lower panel uses exactly three topic papers published in 2010--2022. Article and cell mean regressions include field fixed effects and use HC3 intervals. The final row averages the six separately estimated field slopes and uses a $t_5$ interval. Units are articles, field and year cells, or field slopes, as appropriate.
\end{tablenotes}\end{threeparttable}
\end{table*}

\begin{table*}[!t]
\centering\small
\begin{threeparttable}
\caption{Sensitivity of recent AI intensity contrasts to the topic vocabulary.}
\label{tab:s_ai_definitions}
\begin{tabular}{@{}llrrrr@{}}
\toprule
Outcome & Definition & Estimate & 95\% CI & $p$ & $N$ \\
\midrule
Paper topic breadth & Broad & $0.000000$ & $[-0.004670, 0.004671]$ & 1.000 & 16,591 \\
 & Legacy core & $-0.000004$ & $[-0.004688, 0.004679]$ & 0.999 & 16,591 \\
\addlinespace[2pt]
$\log(1+\mathrm{team\ size})$ & Broad & $0.0127$ & $[-0.0166, 0.0419]$ & 0.396 & 20,393 \\
 & Legacy core & $0.0128$ & $[-0.0163, 0.0418]$ & 0.389 & 20,393 \\
\addlinespace[2pt]
Cited knowledge breadth & Broad & $0.0047$ & $[-0.0127, 0.0220]$ & 0.596 & 1,360 \\
 & Legacy core & $0.0053$ & $[-0.0116, 0.0222]$ & 0.539 & 1,360 \\
\addlinespace[2pt]
Individual prior breadth & Broad & $0.0235$ & $[-0.0005, 0.0476]$ & 0.055 & 661 \\
 & Legacy core & $0.0235$ & $[-0.0006, 0.0476]$ & 0.056 & 661 \\
\addlinespace[2pt]
Individual focal depth & Broad & $-0.0467$ & $[-0.1719, 0.0785]$ & 0.465 & 674 \\
 & Legacy core & $-0.0478$ & $[-0.1720, 0.0764]$ & 0.451 & 674 \\
\bottomrule\end{tabular}
\begin{tablenotes}\footnotesize\item Entries are the post-2022 interaction with baseline AI intensity, standardized within field. All specifications use the same 2010--2018 exposure window, subfield and year fixed effects, and standard errors clustered by subfield. The paper, team, and cited knowledge analyses retain 83 subfields; each contributor outcome retains 77. Confidence intervals and $p$ values are unadjusted exploratory summaries. The legacy core and broad vocabularies are defined in the accompanying text.
\end{tablenotes}\end{threeparttable}
\end{table*}

\subsection{Team selection and aggregated trends}

Table~\ref{tab:s_selection} reports the prevalence of papers above the contributor sampling threshold. Across the six fields, the share with more than 20 recorded authors rose from 0.41\% in 2010--2014 to 0.97\% in 2023--2025. The recent share was largest in physics and astronomy (3.80\%) and medicine (1.80\%), locating the main change in sample coverage in those fields. Restricting the exactly three topic analysis to teams of 1--20 authors retained 29,329 pre-2023 papers and an annual slope of --0.00122 [--0.00169, --0.00076]. The output concentration trend therefore also characterizes articles within the contributor sampling frame.

We further averaged breadth within each field and year over 2010--2022 and regressed the 78 cell means on year and field fixed effects, giving each cell equal weight. The annual slope was --0.00111 [--0.00184, --0.00038] under HC3 errors ($p=0.003$). This aggregate analysis preserves the negative pooled trajectory with the field and year cell as the observation unit. A complementary summary first estimated a separate slope in each field, then used the mean and standard error of the six slopes with a $t_5$ interval. It gave the same mean of --0.00111 but a wider interval [--0.00266, 0.00044] ($p=0.125$). The wider interval reflects between-field variation, distinguishing the average trend within the sampled cells from its generalization across fields.

\subsection{AI vocabulary sensitivity}

The legacy core definition uses six topic name strings, namely artificial intelligence, machine learning, deep learning, neural network, natural language processing, and computer vision. The broad definition adds generative adversarial, large language model, foundation model, and transformer model. Both use case insensitive substring matching. This distinction defines a narrower operational vocabulary; the added terms are not all inventions after 2018. All topic assignments remain drawn from the same current OpenAlex snapshot, so the comparison tests sensitivity to the vocabulary while holding the retrospective classification source fixed.

We retained the same 2010--2018 baseline, Jeffreys smoothing, within-field standardization, subfield eligibility rules, 2019--2025 estimation window, fixed effects, and clustered inference. The broad and legacy definitions marked 431 and 421 baseline papers as AI related. The 10 changed papers arose from the topic Generative Adversarial Networks and Image Synthesis, and the resulting exposures correlated at 0.9996 across the 83 retained subfields. As shown in Table~\ref{tab:s_ai_definitions}, paper breadth remained centered essentially at zero, and team size and cited breadth retained intervals spanning zero under both definitions. Individual breadth and depth also showed closely matching coefficients and intervals. The vocabulary comparison therefore supports a stable characterization of recent heterogeneity by prior AI proximity.

\FloatBarrier

\printbibliography

\end{document}